\documentclass[12pt]{article}
\usepackage{epsf}
\title{ {\bf
The effect of the location of the new Higgs doublet on the
radiative lepton flavor violating decays in the split fermion
scenario.}}
\author{\vspace{1cm}\\
        {\bf E. O. Iltan}
        \thanks{E-mail address:
        eiltan@heraklit.physics.metu.edu.tr}
 \\
        Physics Department, Middle East Technical University \\
        Ankara, Turkey\\}

\date{}

\begin{document}
\setlength{\baselineskip}{24pt}
\maketitle
\setlength{\baselineskip}{7mm}
\begin{abstract}
We study the branching ratios of the lepton flavor violating
processes $\mu\rightarrow e\gamma$, $\tau\rightarrow e\gamma$  and
$\tau\rightarrow \mu\gamma$ in the split fermion scenario, with
the assumption that the new Higgs doublet is restricted to the 4D
brane (thin bulk) in one and two extra dimensions, in the
framework of the two Higgs doublet model. We observe that the
branching ratios are sensitive to the location of the 4D brane
and, in the second case, the width of the thin bulk, especially
for the $\mu\rightarrow e \gamma$ decay.
\end{abstract}
\thispagestyle{empty}
\newpage
\setcounter{page}{1}
\section{Introduction}
Since the existence of the radiative lepton flavor violating (LFV)
decays depends on the flavor changing currents, they appear in the
loop level and, therefore, they are rich from the theoretical
point of view. In the standard model (SM), even with the neutrino
mixing with non zero neutrino masses, their calculated branching
ratios (BRs) are too small to reach the experimental limits. This
motivates one to search new models beyond the SM and to extend the
particle spectrum in order to enhance the BRs of these decays.

The discoveries of heavy leptons stimulated the experimental work
for the LFV decays. The current limits for the (BRs) of
$\mu\rightarrow e\gamma$ and $\tau\rightarrow e\gamma$ decays are
$1.2\times 10^{-11}$ \cite{Brooks} and $3.9\times 10^{-7}$
\cite{Hayasaka}, respectively. To search for the LFV decay
$\mu\rightarrow e \gamma$ \cite{Nicolo} a new experiment at PSI
has been described and aimed to reach to a sensitivity of $BR\sim
10^{-14}$, improved by three order of magnitudes with respect to
previous searches. At present the experiment (PSI-R-99-05
Experiment) is still running in the MEG \cite{Yamada}. On the
other hand, the BR of $\tau\rightarrow \mu\gamma$ decay has been
measured as $1.1\times 10^{-6}$ \cite{Ahmed}, and recently, an
upper limit of $BR=9.0\, (6.8)\, 10^{-8}$ at $90\%$ CL has been
obtained \cite{Roney} (\cite{Aubert}), which is an improvement
almost by one order of magnitude with respect to previous one.

Besides the experimental work, there is  an extensive theoretical
analysis done on the radiative LFV decays in the literature
\cite{Barbieri1}-\cite{Paradisi}. These decays are studied in the
supersymmetric models \cite{Barbieri1}, in the framework of the
two Higgs doublet model (2HDM) \cite{Iltan1, Diaz, IltanExtrDim,
IltanLFVSplit} and in a model independent way \cite{Chang}.
Recently, they are analyzed in the framework of 2HDM and the
supersymmetric model in \cite{Paradisi}.

In the present work, we study the LFV processes $\mu\rightarrow
e\gamma$, $\tau\rightarrow e\gamma$  and $\tau\rightarrow
\mu\gamma$ in the 2HDM and we respect the idea that the hierarchy
of fermion masses is coming from the overlap of the fermion
Gaussian profiles in the extra dimensions, so called the split
fermion scenario \cite{Hamed}. Here, the extension of the Higgs
sector and the permission of the flavor changing neutral currents
(FCNC's) at tree level, ensure the BRs of the radiative decays
under consideration to be enhanced, theoretically. In addition to
this, the extension of the space-time, the inclusion of one (two)
extra spatial dimension, causes certain modifications in the BRs.

In the split fermion scenario, the fermions are assumed to locate
at different points in the extra dimensions with the exponentially
small overlaps of their wavefunctions and there are various
studies in the literature \cite{Hamed}-\cite{Delgado}. One of the
phenomenologically reliable set of explicit positions of left and
right handed components of fermions in a single extra dimension
have been predicted in \cite{Mirabelli} and this is the one we use
in our numerical calculations. In \cite{Changg}, the restrictions
on the split fermions in the extra dimensions have been obtained
by using the leptonic W decays and the lepton violating processes,
and the CP violation in the quark sector has been studied in
\cite{Branco}. \cite{Chang2} is devoted to find stringent bounds
on the size of the compactification scale 1/R, the physics of
kaon, neutron and B/D mesons  and, in \cite{Hewett}, the rare
processes in the split fermion scenario have been studied. The
shapes and overlaps of the fermion wave functions in the split
fermion model has been estimated in \cite{Perez} and the work in
\cite{IltanLFVSplit} (\cite{IltanEDMSplit},\cite{IltanZl1l2Split})
is related to the the radiative LFV decays  (electric dipole
moments of charged leptons, the LFV $Z\rightarrow l_i\,l_j$
decays) in the split fermion scenario. Recently, the Higgs
localization in the split fermion models has been studied in
\cite{Surujon}.

In our calculations, we consider that the leptons have Gaussian
profiles in the extra dimension(s). Furthermore, we first assume
that the new Higgs doublet lies in 4D brane, whose coordinate(s)
in one (two) extra dimension(s) is arbitrary,
$y_p\,\sigma\,(y_p\sigma, z_p\sigma)$ where $\sigma$ is the width
of the Gaussian lepton profile in the extra dimension(s). Second,
we take that the new Higgs doublet lies in one (two) extra
dimension(s) but restricted into the thin bulk which has width
$w\,R\,(w_y\,R, w_z\,R)$, $w\leq 2\,\pi$ ($w_y\leq 2\,\pi$,
$w_z\leq 2\,\pi$). We observe that the BRs are sensitive to the
location of the 4D brane and, in the second case, the width of the
thin bulk,  especially for the $\mu\rightarrow e \gamma$ decay.

The paper is organized as follows: In Section 2, we present the
lepton-lepton-new Higgs scalar vertices and the BRs of the
radiative LFV decays in the split fermion scenario with the
assumption that the new Higgs doublet is restricted to the 4D
brane (thin bulk) in the one and two extra dimensions, in the
framework of the 2HDM. Section 3 is devoted to discussion and our
conclusions.
\section{The possible effects on the radiative LFV
decays in the split fermion scenario, due to the different
locations of the new Higgs doublet in the extra dimensions, in the
2HDM }
The radiative LFV decays $l_i\rightarrow l_j\,\gamma$ exist at
least at one loop level in the SM and the numerical values of the
BRs of these decays are extremely small. To enhance them, one goes
beyond the SM and the version of the 2HDM, permitting the
existence of the FCNCs at tree level, is one of the candidate to
obtain relatively large BRs, since the extended Higgs sector
brings additional contributions. These contributions are
controlled by the new Yukawa couplings, which are complex in
general. Besides, the inclusion of the spatial extra dimensions
causes to enhance the BRs, since the particle spectrum is further
extended after the compactification. In our analysis, we consider
effects of the extension of the Higgs sector and the extra
dimensions. Here we respect the split fermion scenario which is
based on the idea that the hierarchy of lepton masses are coming
from the lepton Gaussian profiles in the extra dimensions.

Now, we present the Yukawa Lagrangian responsible for these
interactions in a single (two) extra dimension, respecting the
split fermion scenario,
\begin{eqnarray}
{\cal{L}}_{Y}=
\xi^{E}_{5\,(6) \,ij} \bar{\hat{l}}_{i L} \phi_{2} \hat{E}_{j R} +
h.c. \,\,\, , \label{lagrangian}
\end{eqnarray}
where $L$ and $R$ denote chiral projections $L(R)=1/2(1\mp
\gamma_5)$, $\phi_{2}$ is the new scalar doublet and
$\xi^{E}_{5\,(6)\, ij}$ are the flavor violating complex Yukawa
couplings in five (six) dimensions. Here, $\hat{l}_{i L}$
($\hat{E}_{j R}$), with family indices $i,j$, are the zero
mode\footnote{Notice that we take only the zero mode lepton fields
in our calculations.} lepton doublets (singlets) with Gaussian
profiles in the extra dimension(s) $y$ ($y,z$) and, in a single
extra dimensions, they read
\begin{eqnarray}
\hat{l}_{i L}&=& N\,e^{-(y-y_{i L})^2/2 \sigma^2}\,l_{i L} ,
\nonumber
\\ \hat{E}_{j R}&=&N\, e^{-(y-y_{j R})^2/2 \sigma^2}\, E_{j R}\, ,
\label{gaussianprof}
\end{eqnarray}
with the normalization factor $N=\frac{1}{\pi^{1/4}\,
\sigma^{1/2}}$. In the case that the leptons are accessible two
both dimensions with Gaussian profiles, we get
\begin{eqnarray}
\hat{l}_{i L}&=& N\,e^{-\Big((y-y_{i L})^2+(z-z_{i L})^2\Big)/2
\sigma^2}\,l_{i L} , \nonumber
\\ \hat{E}_{j R}&=&N\, e^{-\Big((y-y_{j R})^2+(z-z_{j R})^2\Big)/2
\sigma^2}\, E_{j R}\, , \label{gaussianprof2}
\end{eqnarray}
with the normalization factor $N=\frac{1}{\pi^{1/2}\, \sigma}$.
Here, $l_{i L}$ ($E_{j R}$) are the lepton doublets (singlets) in
four dimensions and the parameter $\sigma$ represents the Gaussian
width of the leptons, satisfying the property $\sigma << R$, where
$R$ is the compactification radius. In eq. (\ref{gaussianprof})
(eq. (\ref{gaussianprof2})), the parameters $y_{i L}$- $y_{iR}$
($y_{i L}$ and $z_{i L}$ - $y_{iR}$ and $z_{iR}$) are the fixed
positions of the peaks of left-right handed parts of $i^{th}$
lepton in the fifth (fifth and sixth) dimension and they are
obtained by taking the observed lepton masses into account ( see
\cite{Mirabelli} for a single extra dimension case). The
underlying idea is that the mass hierarchy of leptons are coming
from the relative positions of the Gaussian peaks of the wave
functions located in the extra dimension \cite{Hamed, Mirabelli}.
One possible set of locations for the lepton fields in the fifth
dimension read (see \cite{Mirabelli} for details)
\begin{eqnarray}
P_{l_i}=\sqrt{2}\,\sigma\, \left(\begin{array}{c c c}
11.075\\1.0\\0.0
\end{array}\right)\,,\,\,\,\, P_{e_i}=\sqrt{2}\,\sigma\, \left(\begin{array}
{c c c} 5.9475\\4.9475\\-3.1498
\end{array}\right)
 \,\, . \label{location}
\end{eqnarray}
For two extra dimensions, a possible positions of left handed and
right handed leptons can be obtained by using the observed masses
\footnote{The calculation is similar to the one presented in
\cite{Mirabelli} which is done for a single extra dimension.}.
With the assumption that the lepton mass matrix is diagonal, one
of the possible set of locations for the Gaussian peaks of the
lepton fields in the two extra dimensions are \cite{IltanEDMSplit}
\begin{eqnarray}
P_{l_i}=\sqrt{2}\,\sigma\, \left(\begin{array}{c c c}
(8.417,8.417)\\(1.0,1.0)\\(0.0,0.0)
\end{array}\right)\,,\,\,\,\,
P_{e_i}=\sqrt{2}\,\sigma\, \left(\begin{array} {c c c}
(4.7913,4.7913)\\(3.7913,3.7913)\\(-2.2272,-2.2272)
\end{array}\right)
 \,\, , \label{location2}
\end{eqnarray}
where the numbers in the parenthesis denote the $y$ and $z$
coordinates of the location of the Gaussian peaks of lepton
flavors in the extra dimensions. Notice that we choose the same
numbers for the $y$ and $z$ locations of the Gaussian peaks.

Here, we take that the new Higgs sector does not mix with the old
one and collect SM (new) particles in the first (second) doublet.
We choose the Higgs doublets $\phi_{1}$ and $\phi_{2}$ as
\begin{eqnarray}
\phi_{1}=\frac{1}{\sqrt{2}}\left[\left(\begin{array}{c c}
0\\v+H^{0}\end{array}\right)\; + \left(\begin{array}{c c} \sqrt{2}
\chi^{+}\\ i \chi^{0}\end{array}\right) \right]\, ;
\phi_{2}=\frac{1}{\sqrt{2}}\left(\begin{array}{c c} \sqrt{2}
H^{+}\\ H_1+i H_2 \end{array}\right) \,\, , \label{choice}
\end{eqnarray}
with the vacuum expectation values,
\begin{eqnarray}
<\phi_{1}>=\frac{1}{\sqrt{2}}\left(\begin{array}{c c}
0\\v\end{array}\right) \,  \, ; <\phi_{2}>=0 \,\, ,\label{choice2}
\end{eqnarray}
and $H_1$ and $H_2$ are the mass eigenstates $h^0$ and $A^0$
respectively since no mixing occurs between two CP-even neutral
bosons $H^0$ and $h^0$ at tree level. In this case, the first
Higgs doublet is responsible for the hierarchy of lepton masses
and the LFV interaction at tree level is carried by the new Higgs
field $\phi_{2}$. Now, we follow two possibilities:

\begin{itemize}
\item The new Higgs doublet lies in 4D brane, whose coordinate(s)
in one (two) extra dimension(s) is arbitrary,
$y_p\,\sigma\,(y_p\,\sigma, z_p\,\sigma)$ and leptons are living
in one (two) extra dimension(s);
\item The new Higgs doublet lies in one (two) extra dimension(s)
but restricted into the thin bulk which has width $w\,R\,(w_y\,R,
w_z\,R)$, $w\leq 2\,\pi$ ($w_y\leq 2\,\pi$, $w_z\leq 2\,\pi$) and
leptons are living in one (two) extra dimension;
%
\end{itemize}
and present the lepton-lepton-$S$, $S=h^0,A^0$ vertex factors
after the integration over the extra dimension(s).

\subsection{The vertex factors for the case that new Higgs doublet
lives in the 4D brane }
The LFV processes $l_i\rightarrow l_j\gamma$ are carried by the
lepton-lepton-$S\,(S=h^0,A^0)$ vertices.  The vertex factors
$V_{LR\,(RL)\,ij}$ in the vertices $\bar{\hat{f}}_{iL\, (R)}\,S
(y_p)\, \hat{f}_{j R\, (L)}$ with the right (left) handed $i^{th}$
flavor lepton fields $\hat{f}_{j R\, (L)}$ (see eq.
(\ref{gaussianprof})) and the new Higgs bosons $S$ on the 4D brane
with the location $y=y_p\,\sigma$ in five dimensions , are
obtained by the integration over the fifth dimension and they read
\begin{eqnarray}
V^{br}_{LR\,(RL)\,ij}=V^{br,0}_{LR\,(RL)\,ij}\,V^{br,
extr}_{LR\,(RL)\,ij}\, \,\, , \label{Vbrij11}
\end{eqnarray}
where
\begin{eqnarray}
V^{br, extr}_{LR\,(RL)\,ij}=
 e^{-y_p \,(-y_p \, \sigma
+y_{i L\, (R)}+y_{j R\, (L)})/\sigma} \,\, . \label{Vbrij1}
\end{eqnarray}
Here the factor $V^{br,0}_{LR\,(RL)\,ij}$ is included in the
definition of the coupling in four dimensions

\begin{eqnarray}
\xi^{E}_{ij}\,\Big((\xi^{E \dagger}_{ij})^\dagger\Big)=
V^{br,0}_{LR\,(RL)\,ij} \, \xi^{E}_{5\, ij}\,\Big((\xi^{E}_{5\,
ij})^\dagger\Big)/2 \pi R \,\, , \label{coupbrl4}
\end{eqnarray}
and it is obtained as
\begin{eqnarray}
V^{br,0}_{LR\,(RL)\,ij}=\frac{2\,\sqrt{\pi}\,
R}{\sigma}\,e^{-(y_{i L\, (R)}^2+y_{j R\, (L)}^2)/2\,\sigma^2}
\,\, , \label{V0brij1}
\end{eqnarray}
by choosing that the new Higgs bosons place on the 4D brane at
$y=0$. In the case of two extra dimensions where the leptons feel,
the lepton-lepton-$S$ vertex factors $V^{br, (2)}_{LR\,(RL)\,ij}$
in the vertices $\bar{\hat{f}}_{iL\, (R)}\,S(y_p,z_p)\, \hat{f}_{j
R\, (L)}$, with the right (left) handed $i^{th}$ flavor lepton
fields $\hat{f}_{j R\, (L)}$ (see eq. (\ref{gaussianprof2})) and
the Higgs bosons $S$ restricted on the brane at $y=y_p\,\sigma,
z=z_p\,\sigma$ in six dimensions, are obtained by the integration
over the fifth and sixth dimensions as
\begin{eqnarray}
V^{br,(2)}_{LR\,(RL)\,ij}=V^{br,0,\,(2)}_{LR\,(RL)\,ij}\,
V^{br,\,extr2}_{LR\,(RL)\,ij}\,\, , \label{Vbrij22}
\end{eqnarray}
where
\begin{eqnarray}
V^{br,\,extr2}_{LR\,(RL)\,ij}=\mathbf{e}^{\Big(y_p \,(-y_p \,
\sigma +y_{i L\, (R)}+y_{j R\, (L)})+z_p \,(-z_p \, \sigma +z_{i
L\, (R)}+z_{j R\, (L)})\Big)/\sigma} \,\, . \label{Vbrij2}
\end{eqnarray}
Here the factor $V^{br,0\,(2)}_{LR\,(RL)\,ij}$ is included in the
definition of the coupling in four dimensions
\begin{eqnarray}
\xi^{E}_{ij}\,\Big((\xi^{E}_{ij})^\dagger\Big)=
V^{br,0\,(2)}_{LR\,(RL)\,ij} \, \xi^{E}_{6 \,ij}\,\Big((\xi^{E}_{6
\,ij})^\dagger\Big)/(2 \pi R)^2\, , \label{coupbr2l44}
\end{eqnarray}
and it is obtained by choosing that the new Higgs bosons place in
the 4D brane at $y,z=0$:
\begin{eqnarray}
V^{br,0,\,(2)}_{LR\,(RL)\,ij}=\frac{4\,\pi\,
R^2}{\sigma^2}\,e^{-(y_{i L\, (R)}^2+y_{j R\, (L)}^2+z_{i L\,
(R)}^2+z_{j R\,(L)}^2)/2\,\sigma^2}  \,\, . \label{V0brij2}
\end{eqnarray}
\subsection{The vertex factors in the case that the new Higgs doublet lies
in the thin bulk placed around the origin, in the bulk}
If we consider that the new Higgs doublet lies in a single extra
dimension and is restricted into the thin bulk, having a width
$w\,R$, $w\leq 2\,\pi$, which is placed symmetrically around the
origin, after the compactification on the orbifold $S^1/Z_2$, it
reads
\begin{eqnarray}
\phi_{2}(x,y ) & = & {1 \over {\sqrt{w\, R}}} \left\{
\phi_{2}^{(0)}(x) + \sqrt{2} \sum_{n=1}^{\infty} \phi_{2}^{(n)}(x)
\cos\,(\frac{2\,\pi\,n\,y}{w\,R})\right\} \,,
\label{SecHiggsField}
\end{eqnarray}
where $\phi_{2}^{(0)}(x)$ ($\phi_{2}^{(n)}(x)$) is  the Higgs
doublet in the four dimensions (the KK modes) including the
charged Higgs boson $H^+$ ($H^{(n)+}$), the neutral CP even-odd
Higgs bosons $h^0$- $A^0$ ($h^{0 (n)}$- $A^{0 (n)}$). The non-zero
$n^{th}$ KK mode of the charged Higgs mass is
$\sqrt{m_{H^\pm}^2+m_n^2}$, and the neutral CP even (odd) Higgs
mass is $\sqrt{m_{h^0}^2+m_n^2}$, ($\sqrt{m_{A^0}^2+m_n^2}$ ),
with the $n$'th level KK particle mass
$m_n=\frac{2\,\pi\,n}{w\,R}$. In the two extra dimensions, after
the compactification on the orbifold $(S^1\times S^1)/Z_2$, the
new Higgs field $\phi_{2}$ is expanded as
\begin{eqnarray}
\phi_{2}(x,y,z ) & = & {1 \over {\sqrt{w_y\,w_z}\, R}} \left\{
\phi_{2}^{(0,0)}(x) + 2 \sum_{n,s}^{\infty} \phi_{2}^{(n,s)}(x)
\cos\,\Big(\frac{2\,\pi}{R}\,(\frac{n y}{w_y}+\frac{s
z}{w_z})\Big)\right\} \,. \label{SecHiggsField2}
\end{eqnarray}
where $w_y\,R$ and $w_z\,R$ are widths of thin rectangular bulk
volume, having the center at the origin, with, $w_y\leq 2\,\pi$
and $w_z\leq 2\,\pi$. The KK modes of charged (neutral CP even,
neutral CP odd) Higgs fields existing in the new Higgs doublet
have the masses $\sqrt{m_{H^\pm}^2+m_n^2+m_s^2}$,
($\sqrt{m_{h^0}^2+m_n^2+m_s^2}$, $\sqrt{m_{A^0}^2+m_n^2+m_s^2}$ )
where $m_{n\,(s)}=\frac{2\,\pi\,n\,(s)} {w_{y\,(z)\,R}}$  are the
masses of $n\,(s)$'th level KK modes.
Now, we present the lepton-lepton-$S$ vertex factors
$V^{bulk}_{LR\,(RL)\,ij}$ in the vertices $\bar{\hat{f}}_{iL\,
(R)}\,S^{(0)}(x,y)\, \hat{f}_{j R\, (L)}$, with $S^{(0)}=h^0,A^0$
zero modes and the right (left) handed $i^{th}$ flavor lepton
fields $\hat{f}_{j R\, (L)}$ in five dimensions (see eq.
(\ref{gaussianprof})). After the integration over the fifth
dimension we get
\begin{eqnarray}
V^{bulk}_{LR\,(RL)\,ij}=V^{bulk,0}_{LR\,(RL)\,ij}\,
V^{bulk,extr}_{LR\,(RL)\,ij}\,\, , \label{Vbuij1a}
\end{eqnarray}
where
\begin{eqnarray}
V^{bulk,extr}_{LR\,(RL)\,ij}=
\sqrt{\frac{\pi}{2\,w}}\,\Big(Erf[f^{(0)}_{LR\,(RL)\,ij}(y_p)]-
Erf[f^{(w)}_{LR\,(RL)\,ij}(y_p)]\Big) \,\, , \label{Vbuij1exa}
\end{eqnarray}
with
\begin{eqnarray}
f^{(w)}_{LR\,(RL)\,ij}(y_p)&=&\frac{-2\,(y_p+w)\,R+y_{i L\,
(R)}+y_{j R\,(L)}}{2\,\sigma}\,\, ,\nonumber \\
f^{(0)}_{LR\,(RL)\,ij}(y_p)&=&f^{(w)}_{LR\,(RL)\,ij}(y_p)|_{w\rightarrow
0} \,\, , \label{fij1a}
\end{eqnarray}
and $y_p=-w/2$.  Here the function $Erf[z]$ is the so called error
function and it is defined as
\begin{eqnarray}
Erf[z]=\frac{2}{\sqrt{\pi}}\,\int_{0}^{z}\,e^{-t^2}\,dt \,\, .
\label{erffunc}
\end{eqnarray}
Notice that the factor $V^{bulk,0}_{LR\,(RL)\,ij}$ is included in
the definition of the coupling in four dimensions as
\begin{eqnarray}
\xi^{E}_{ij}\,\Big((\xi^{E}_{ij})^\dagger\Big)=
V^{bulk,0}_{LR\,(RL)\,ij} \, \xi^{E}_{5\, ij}\,\Big((\xi^{E}_{5\,
ij})^\dagger\Big)/\sqrt{2 \pi R} \,\, , \label{coupbul4}
\end{eqnarray}
where
\begin{eqnarray}
V^{bulk,0}_{LR\,(RL)\,ij}=e^{-(y_{i L\, (R)}-y_{j R\, (L)})^2/4
\sigma^2} \,\, , \label{V0buij1}
\end{eqnarray}
which is responsible for the hierarchy of lepton masses and it is
the only factor appearing in the case that the size of the thin
bulk, restricting the new Higgs bosons, is $2\pi R$ (This is the
case that the factor $V^{bulk,extr}_{LR\,(RL)\,ij}$ is unity.). In
the case of the interaction of the KK modes for the thin bulk of S
boson with the leptons, the lepton-lepton-$S^{(n)}$ vertex factor
$V^{bulk,n}_{LR\,(RL)\,ij}$ in the vertices $\bar{\hat{f}}_{iL\,
(R)}\,S^{(n)}(x,y)\, \hat{f}_{j R\, (L)}$ is obtained by the
integration over the fifth dimension and it reads
\begin{eqnarray}
V^{bulk,n}_{LR\,(RL)\,ij}= V^{bulk,0}_{LR\,(RL)\,ij}
\,V^{bulk,(n),\, extr}_{LR\,(RL)\,ij} \,\, , \label{Vbunij1a}
\end{eqnarray}
where
\begin{eqnarray}
V^{bulk,(n),extr}_{LR\,(RL)\,ij}&=&\frac{1}{2}\,\sqrt{\frac{\pi}{2\,w}}\,
\Bigg\{\nonumber \\ & & \!\!\!\! e^{+(n)}_{LR\,(RL)\,ij}\,
\Big(Erf[f^{(0)}_{LR\,(RL)\,ij}(y_p)+\frac{i\,n\,\pi\,\rho}{w}]-
Erf[f^{(w)}_{LR\,(RL)\,ij}(y_p)+\frac{i\,n\,\pi\,\rho} {w}]
\Big)\nonumber \\ \!\!&+& \!\!\!\! e^{-(n)}_{LR\,(RL)\,ij}
\Big(Erf[f^{(0)}_{LR\,(RL)\,ij}(y_p)-\frac{i\,n\,\pi\,\rho}{w}]-
Erf[f^{(w)}_{LR\,(RL)\,ij}(y_p)-\frac{i\,n\,\pi\,\rho} {w}]
\Big)\!\! \Bigg\}, \label{Vbunij1exa}
\end{eqnarray}
and
\begin{eqnarray}
e^{\pm (n)}_{LR\,(RL)\,ij}= Exp\Big[-\frac{n^2
\pi^2\,\rho^2}{w^2}\pm \frac{i\,n\,\pi\,((y_{i L\, (R)}+y_{i R\,
(L)}) }{R\,w}\Big] \,\, . \label{e1KK}
\end{eqnarray}
Here the function $f^{(w)}_{LR\,(RL)\,ij}(y_p)$ is defined in eq.
(\ref{fij1a}) and the parameter $\rho$ is defined as
$\rho=\sigma/R$.

If the new Higgs bosons and leptons are accessible to two extra
dimensions, the lepton-lepton-$S$ vertex factors
$V^{bulk,(2)}_{LR\,(RL)\,ij}$ in the vertices $\bar{\hat{f}}_{iL\,
(R)}\,S^{(0,0)}(x,y,z)\, \hat{f}_{j R\, (L)}$, with
$S^{(0,0)}=h^0,A^0$ zero modes and the right (left) handed
$i^{th}$ flavor lepton fields $\hat{f}_{j R\, (L)}$ in six
dimensions (see eq. (\ref{gaussianprof2})), are obtained by the
integration over the fifth and sixth dimensions and as
\begin{eqnarray}
V^{bulk,(2)}_{LR\,(RL)\,ij}=V^{0,0}_{LR\,(RL)\,ij}\,V^{bulk,\,
extr2}_{LR\,(RL)\,ij}\,\, , \label{Vbuij1a22}
\end{eqnarray}
where
\begin{eqnarray}
V^{bulk,\, extr2}_{LR\,(RL)\,ij}&=& \frac{\pi}{2}
\,\sqrt{\frac{1}{w_y\,w_z}}\,
\prod_{r=y,z}\, \Big(Erf[f^{(0)}_{LR\,(RL)\,ij}(r_p)]\nonumber \\
&-& Erf[f^{(w_r)}_{LR\,(RL)\,ij}(r_p)]\Big) \,\, .
\label{Vbuij1a2}
\end{eqnarray}
Notice that the factor $V^{bulk,\, 0,0}_{LR\,(RL)\,ij}$ is
included in the definition of the coupling in four dimensions:
\begin{eqnarray}
\xi^{E}_{ij}\,\Big((\xi^{E}_{ij})^\dagger\Big)= V^{bulk, \,
0,0}_{LR\,(RL)\,ij} \, \xi^{E}_{6 \,ij}\,\Big((\xi^{E}_{6
\,ij})^\dagger\Big)/(2 \pi R)\, , \label{coupbul44a2}
\end{eqnarray}
where
\begin{eqnarray}
V^{bulk,\, 0,0}_{LR\,(RL)\,ij}=e^{-\Big((y_{i L\, (R)}-y_{i R\,
(L)})^2+(z_{i L\, (R)}-z_{i R\, (L)})^2\Big)/4 \sigma^2} \,\, ,
\label{V0buij2a}
\end{eqnarray}
which is responsible for the hierarchy of lepton masses in the
case of two extra dimensions and it is the only factor appearing
when the size of the thin bulk, restricting the new Higgs bosons,
is $2\pi R$ in both directions (This is the case that the factor
$V^{bulk,\,extr2}_{LR\,(RL)\,ij}$ is unity.). On the other hand,
the lepton-lepton-$S^{(n,s)}$ vertex factor $V^{bulk,
n,s}_{LR\,(RL)\,ij}$ in the vertices $\bar{\hat{f}}_{iL\,
(R)}\,S^{(n,s)}(x,y)\, \hat{f}_{j R\, (L)}$ is obtained by the
integration over the fifth and six dimensions and it reads
\begin{eqnarray}
V^{bulk, n,s}_{LR\,(RL)\,ij}= V^{bulk,\, 0,0}_{LR\,(RL)\,ij}
\,V^{bulk, (n,s),\, extr2}_{LR\,(RL)\,ij} \,\, ,
\label{Vbuij1a22ex}
\end{eqnarray}
with
\begin{eqnarray}
V^{bulk,(n,s),\,extr2}_{LR\,(RL)\,ij}&=&\frac{\pi}{4}\,\sqrt{\frac{1}{w_y\,w_z}}
\Bigg\{ \nonumber \\ & & e^{+(n,s)}_{LR\,(RL)\,ij}\,
\prod_{(r;t)=^{(y;n)}_{(z;s)} } \,
\Big(Erf[f^{(0)}_{LR\,(RL)\,ij}(r_p)+\frac{i\,t\,\pi\,\rho}{w_r}]
\nonumber \\
&-& Erf[f^{(w_r)}_{LR\,(RL)\,ij}(r_p)+\frac{i\,t\,\pi\,\rho}
{w_r}]\Big)\nonumber \\
&+& e^{-(n,m)}_{LR\,(RL)\,ij}\, \prod_{(r;t)=^{(y;n)}_{(z;s)}}\,
\Big(Erf[f^{(0)}_{LR\,(RL)\,ij}(r_p)-\frac{i\,t\,\pi\,\rho}
{w_r}]\nonumber \\
&-& Erf[f^{(w_r)}_{LR\,(RL)\,ij}(r_p)-\frac{i\,t\,\pi\,\rho}
{w_r}]\Big) \Bigg\} \,\, , \label{Vbuij1a2ex}
\end{eqnarray}
where
\begin{eqnarray}
e^{\pm (n,s)}_{LR\,(RL)\,ij}=
Exp\Big[-\pi^2\,\rho^2\,(\frac{n^2}{w_y^2}+\frac{s^2}{w_z^2}) \pm
\frac{i\,\pi}{R}\Big( \frac{n\,((y_{i L\, (R)}+y_{i R\ (L)})
}{w_y}+\frac{s\,((z_{i L\, (R)}+z_{i R\ (L)}) }{w_z}\Big)\Big]
\,\, . \label{e1KKbulk}
\end{eqnarray}
\subsection{The decay widths of LFV decays $l_i\rightarrow l_j\gamma$}
\begin{figure}[htb]
\vskip 2.0truein \centering \epsfxsize=6.8in
\leavevmode\epsffile{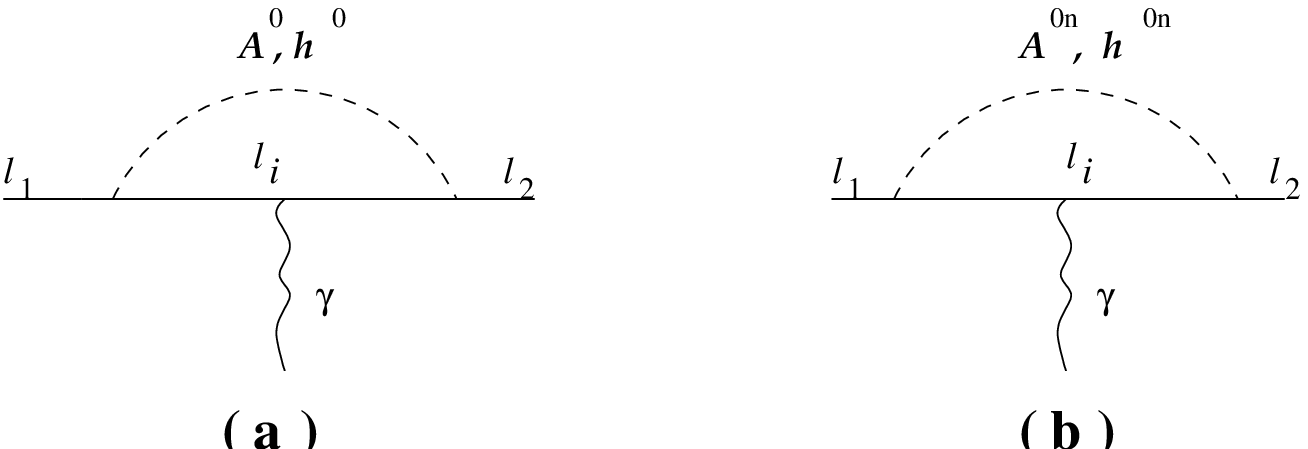} \vskip 1.0truein \caption[]{One loop
diagrams contribute to $l_1\rightarrow l_2 \gamma$ decay  due to
the zero mode (KK mode) neutral Higgs bosons $h^0$ and $A^0$
($h^{0 n}$ and $A^{0 n}$) in the 2HDM, for a single extra
dimension. Here $l_i$ represents the internal charged lepton.}
\label{fig1}
\end{figure}

Now, we will present the decay widths of the LFV processes
$\mu\rightarrow e\gamma$, $\tau\rightarrow e\gamma$ and
$\tau\rightarrow \mu\gamma$ for both cases. Since they exist at
loop level, the logarithmic divergences appear in the calculations
and we eliminate them by using the on-shell renormalization scheme
\footnote{In this scheme, the self energy diagrams for on-shell
leptons vanish since they can be written as $
\sum(p)=(\hat{p}-m_{l_1})\bar{\sum}(p) (\hat{p}-m_{l_2})\, , $
however, the vertex diagrams (see Fig.\ref{fig1}) give non-zero
contribution. In this case, the divergences can be eliminated by
introducing a counter term $V^{C}_{\mu}$ with the relation
$V^{Ren}_{\mu}=V^{0}_{\mu}+V^{C}_{\mu} \, , $ where
$V^{Ren}_{\mu}$ ($V^{0}_{\mu}$) is the renormalized (bare) vertex
and by using the gauge invariance: $k^{\mu} V^{Ren}_{\mu}=0$.
Here, $k^\mu$ is the four momentum vector of the outgoing
photon.}. Taking only $\tau$ lepton for the internal line,
\footnote{We take into account only the internal $\tau$-lepton
contribution since, we respect the Sher scenerio \cite{Sher},
results in the couplings $\bar{\xi}^{E}_{N, ij}$ ($i,j=e,\mu$),
are small compared to $\bar{\xi}^{E}_{N,\tau\, i}$
$(i=e,\mu,\tau)$, due to the possible proportionality of them to
the masses of leptons under consideration in the vertices. Here,
we use the dimensionful coupling $\bar{\xi}^{E}_{N,ij}$ with the
definition $\xi^{E}_{N,ij}=\sqrt{\frac{4\, G_F}{\sqrt{2}}}\,
\bar{\xi}^{E}_{N,ij}$ where N denotes the word "neutral".} the
decay width $\Gamma$ for the $l_i\rightarrow l_j\gamma$ decay
reads
\begin{eqnarray}
\Gamma (l_i\rightarrow l_j\gamma)=c_1(|A_1|^2+|A_2|^2)\,\, ,
\label{DWmuegam}
\end{eqnarray}
for $l_i\,(l_j)=\tau;\mu\,(\mu$ or $e; e)$. Here $c_1=\frac{G_F^2
\alpha_{em} m^3_{l_i}}{32 \pi^4}$, $A_1$ ($A_2$) is the left
(right) chiral amplitude.

If the new Higgs doublet lies in 4D brane, with the coordinate
$y_p\,\sigma$,in a single extra dimension, the amplitudes read,
\begin{eqnarray}
A_1&=&Q_{\tau} \frac{1}{48\,m_{\tau}^2} \Bigg (6\,m_\tau\,
\bar{\xi}^{E *}_{N,\tau f_2}\, \bar{\xi}^{E *}_{N,f_1\tau}\,
V^{br, extr}_{LR\,f_1\,\tau} \, V^{br,
extr}_{RL\,f_2\,\tau}\,\Big( F (v_{h^0})-F (v_{A^0})\Big )
\nonumber \\ &+& m_{f_1}\,\bar{\xi}^{E *}_{N,\tau f_2}\,
\bar{\xi}^{E}_{N,\tau f_1}\, V^{br, extr}_{RL\,f_1\,\tau}\,V^{br,
extr}_{RL\,f_2\,\tau}\, \Big( G (v_{h^0})+G (v_{A^0})\Big) \Bigg)
\nonumber \,\, , \\
A_2&=&Q_{\tau} \frac{1}{48\,m_{\tau}^2} \Bigg (6\,m_\tau\,
\bar{\xi}^{E}_{N, f_2 \tau}\, \bar{\xi}^{E}_{N,\tau f_1}\,
\,V^{br, extr}_{LR\,f_2\,\tau} \, V^{br, extr}_{RL\,f_1\,\tau}\,
\Big(F (v_{h^0})-F (v_{A^0})\Big )\nonumber \\ &+&
m_{f_1}\,\bar{\xi}^{E}_{N,f_2\tau}\, \bar{\xi}^{E *}_{N,f_1
\tau}\,  V^{br, extr}_{LR\,f_2\,\tau} \,V^{br,
extr}_{LR\,f_1\,\tau} \, \Big( G (v_{h^0})+ G (v_{A^0}) \Big)
\Bigg)
 \,\, , \label{A1A2}
\end{eqnarray}
where $v_{S}=\frac{m^2_{\tau}}{m^2_{S}}$, and  $V^{br,
extr}_{LR\,(RL)\,f_1\,\tau}$ and $V^{br,
extr}_{RL\,(LR)\,f_2\,\tau}$ are defined in eq. (\ref{Vbrij1}).
Here the functions $F (w)$ and $G (w)$ are  given by
\begin{eqnarray}
F (w)&=&\frac{w\,(3-4\,w+w^2+2\,ln\,w)}{(-1+w)^3} \, , \nonumber \\
G (w)&=&\frac{w\,(2+3\,w-6\,w^2+w^3+ 6\,w\,ln\,w)}{(-1+w)^4} \,\,
. \label{functions2}
\end{eqnarray}

In the case that new Higgs doublet lies in 4D brane, with the
location $y_p\,\sigma, z_p\,\sigma$ and the leptons have gaussian
profiles in the two extra dimensions, the amplitudes would be the
same except the vertex  factors $V^{br, extr}_{LR\,f_1\,\tau}$ and
$V^{br, extr}_{RL\,f_2\,\tau}$ are replaced by
$V^{br,\,extr2}_{LR\,f_1\,\tau}$ and
$V^{br,\,extr2}_{RL\,f_2\,\tau}$ (see eq. (\ref{Vbrij2})).

As another possibility, we consider that the new Higgs doublet
lies in the one extra dimension but restricted into the thin bulk
which has width $w\,R$, $w\leq 2\,\pi$ and leptons have gaussian
profiles in one extra dimension. Respecting this scenario, the
amplitudes read,
\begin{eqnarray}
A_1&=&Q_{\tau} \frac{1}{48\,m_{\tau}^2} \Bigg \{ 6\,m_\tau\,
\bar{\xi}^{E *}_{N,\tau f_2}\, \bar{\xi}^{E *}_{N,f_1\tau}\,\Bigg(
V^{bulk, extr}_{LR\,f_1\,\tau} \, V^{bulk, extr}_{RL\,f_2\,\tau}\,
\Big( F (v_{h^0})-F (v_{A^0})\Big) \nonumber \\ &+& 2\,
\sum_{n=1}^{\infty}\, V^{bulk,(n),extr}_{LR\,f_1\,\tau} \,
V^{bulk,n,extr}_{RL\,f_2\,\tau} \Big( F (v_{n, h^0})-F (v_{n,
A^0})\Big ) \Bigg ) \nonumber \\ &+& m_{f_1}\,\bar{\xi}^{E
*}_{N,\tau f_2}\, \bar{\xi}^{E}_{N,\tau f_1}\, \Bigg( V^{bulk,
extr}_{RL\,f_1\,\tau}\,V^{bulk,
extr}_{RL\,f_2\,\tau}\, \Big( G (v_{h^0})+G(v_{A^0}) \Big )\nonumber \\
&+& 2\,\sum_{n=1}^{\infty}\, V^{bulk,(n),
extr}_{RL\,f_1\,\tau}\,V^{bulk,n,extr}_{RL\,f_2\,\tau}\,\, \Big( G
(v_{n, h^0})+G (v_{n, A^0})\Big)\Bigg) \Bigg \}
\nonumber \,\, , \\
A_2&=&Q_{\tau} \frac{1}{48\,m_{\tau}^2} \Bigg \{ 6\,m_\tau\,
\bar{\xi}^{E}_{N, f_2 \tau}\, \bar{\xi}^{E}_{N,\tau f_1}\,\Bigg(
V^{bulk, extr}_{LR\,f_2\,\tau} \, V^{bulk, extr}_{RL\,f_1\,\tau}\,
\Big(F(v_{h^0})-F(v_{A^0})\Big)\nonumber \\
&+& 2\,\sum_{n=1}^{\infty}\,V^{bulk,(n), extr}_{LR\,f_2\,\tau} \,
V^{bulk,(n),extr}_{RL\,f_1\,\tau}\,\Big( F (v_{n, h^0})-F (v_{n,
A^0})\Big) \Bigg)\nonumber \\
&+& m_{f_1}\,\bar{\xi}^{E}_{N,f_2\tau}\, \bar{\xi}^{E *}_{N,f_1
\tau}\,\Bigg( V^{bulk, extr}_{LR\,f_2\,\tau} \,V^{bulk,
extr}_{LR\,f_1\,\tau} \,\Big( G (v_{h^0})+G (v_{A^0})\Big) \nonumber \\
&+& 2\,\sum_{n=1}^{\infty}\,V^{bulk,(n), extr}_{LR\,f_2\,\tau}
\,V^{bulk,(n), extr}_{LR\,f_1\,\tau} \, \Big(G (v_{n, h^0})+ G
(v_{n, A^0})\Big) \Bigg) \Bigg\}
 \,\, , \label{A1A22}
\end{eqnarray}
where $v_{n, S}=\frac{m^2_{\tau}}{m^2_{S}+m_n^2}$,
$m_n=\frac{2\,\pi\,n}{w\,R}$ and $Q_{\tau}$ is the charge of
$\tau$ lepton. Here the vertex factors $V^{bulk,
extr}_{LR\,(RL)\,f_1\,\tau}$ and $V^{bulk,
extr}_{RL\,(LR)\,f_2\,\tau}$
($V^{bulk,(n),extr}_{LR\,(RL)\,f_1\,\tau}$ and $V^{bulk,(n),
extr}_{RL\,(LR)\,f_2\,\tau}$) are defined in eq. (\ref{Vbuij1exa})
(eq. (\ref{Vbunij1exa})).

If the new Higgs doublet lies in the two extra dimensions but
restricted into the thin bulk which has widths $w_y\,R, w_z\,R$,
$w_y\leq 2\,\pi, w_z\leq 2\,\pi$ and leptons have gaussian
profiles in two extra dimensions, the amplitudes become,
\begin{eqnarray}
A_1&=&Q_{\tau} \frac{1}{48\,m_{\tau}^2} \Bigg \{ 6\,m_\tau\,
\bar{\xi}^{E *}_{N,\tau f_2}\, \bar{\xi}^{E *}_{N,f_1\tau}\,\Bigg(
V^{bulk, extr2}_{LR\,f_1\,\tau} \, V^{bulk,
extr2}_{RL\,f_2\,\tau}\, \Big( F (v_{h^0})-F (v_{A^0})\Big)
\nonumber \\ &+& 4\, \sum_{(n,s)}^{\infty}\,
V^{bulk,(n,s),\,extr2}_{LR\,f_1\,\tau} \,
V^{bulk,(n,s),\,extr2}_{RL\,f_2\,\tau} \Big( F (v_{(n,m), h^0})-F
(v_{(n,m), A^0})\Big ) \Bigg ) \nonumber \\ &+&
m_{f_1}\,\bar{\xi}^{E *}_{N,\tau f_2}\, \bar{\xi}^{E}_{N,\tau
f_1}\, \Bigg( V^{bulk, extr2}_{RL\,f_1\,\tau}\,V^{bulk,
extr2}_{RL\,f_2\,\tau}\, \Big( G (v_{h^0})+G(v_{A^0}) \Big )\nonumber \\
&+& 4\,\sum_{(n,s)}^{\infty}\, V^{bulk,(n,s),\,
extr2}_{RL\,f_1\,\tau}\,V^{bulk,(n,s),\,extr2}_{RL\,f_2\,\tau}\,\,
\Big( G (v_{(n,s),h^0})+G (v_{(n,s),A^0})\Big)\Bigg) \Bigg \}
\nonumber \,\, , \\
A_2&=&Q_{\tau} \frac{1}{48\,m_{\tau}^2}  \Bigg \{ 6\,m_\tau\,
\bar{\xi}^{E}_{N, f_2 \tau}\, \bar{\xi}^{E}_{N,\tau f_1}\,\Bigg(
V^{bulk, extr2}_{LR\,f_2\,\tau} \, V^{bulk,
extr2}_{RL\,f_1\,\tau}\, \Big( F(v_{h^0})-F(v_{A^0})\Big)\nonumber \\
&+& 4\,\sum_{(n,s)}^{\infty}\,V^{bulk,(n,s),
extr2}_{LR\,f_2\,\tau} \,
V^{bulk,(n,s),extr2}_{RL\,f_1\,\tau}\Big(F (v_{(n,s), h^0})-F
(v_{(n,s), A^0})\Big) \Bigg) \nonumber \\ &+&
m_{f_1}\,\bar{\xi}^{E}_{N,f_2\tau}\, \bar{\xi}^{E *}_{N,f_1
\tau}\,\Bigg( V^{bulk, extr2}_{LR\,f_2\,\tau} \,V^{bulk,
extr2}_{LR\,f_1\,\tau} \,\Big( G (v_{h^0})+G (v_{A^0}\Big) \nonumber \\
&+& 4\,\sum_{(n,s)}^{\infty}\,V^{bulk,(n,s),
extr2}_{LR\,f_2\,\tau} \,V^{bulk,(n,s), extr2}_{LR\,f_1\,\tau} \,
\Big(G (v_{(n,s), h^0})+ G (v_{(n,s), A^0})\Big) \Bigg) \Bigg\}
 \,\, , \label{A1A222}
\end{eqnarray}
where $v_{(n,s),S}=\frac{m^2_{\tau}}{m^2_{S}+m_n^2+m_s^2}$,
$m_{n\,(s)}=\frac{2\,\pi\,n\,(s)}{w_{y\,(z)\,R}}$ and the vertex
factors $V^{bulk, extr2}_{LR\,(RL)\,f_1\,\tau}$ and $V^{bulk,
extr2}_{RL\,(LR)\,f_2\,\tau}$
($V^{bulk,(n,s),extr2}_{LR\,(RL)\,f_1\,\tau}$ and $V^{bulk,(n,s),
extr2}_{RL\,(LR)\,f_2\,\tau}$) are defined in eq. (\ref{Vbuij1a2})
(eq. (\ref{Vbuij1a2ex})). In eq. (\ref{A1A222}) the summation
would be done over $n,s=0,1,2 ...$, except $n=s=0$.
\section{Discussion}
The radiative LFV decays $l_i\rightarrow l_j\gamma$ are based on
the FCNC currents and such currents are permitted at tree level in
the framework of the general 2HDM. Since these decays exist at
least in the one loop of level, the theoretical expressions of the
physical parameters like the BR contains number of free parameters
belonging to the model used and the Yukawa couplings,
$\xi^E_{N,ij}, \, i,j=e, \mu, \tau$, which are strengths of
lepton-lepton-$S\,(S=h^0, A^0)$ vertices, playing essential role
on the physical parameters, are among them. In the present work,
we follow the split fermion scenario, which is among the possible
solutions of the hierarchy of lepton masses, with the assumption
that the lepton Gaussian profiles have different locations in the
extra dimension. Here, we consider two possibilities: The first
one is that the new Higgs doublet lies in 4D brane, whose
coordinate(s) in one (two) extra dimension(s) is arbitrary,
$y_p\,\sigma\,(y_p\,\sigma,z_p\,\sigma)$ and the split leptons are
living in one (two) extra dimension(s). As a second case, we take
that the new Higgs doublet lies in one (two) extra dimension(s)
but restricted into the thin bulk, which has width,
$w\,R\,(w_y\,R\,,w_z\,R\,)$, $w\leq 2\,\pi$ ($w_y\leq 2\,\pi$,
$w_z\leq 2\,\pi$) and leptons are living in one (two) extra
dimension. To get the interaction vertices
lepton-lepton-$S\,(S=h^0, A^0)$ in four dimensions, we make the
integration over extra dimensions(s) and we calculate the
additional factors to the Yukawa couplings. In the case that the
new Higgs doublet lies in 4D brane, we take the additional effects
$V^{br,extr}_{LR\,(RL)\,ij}$ (see eq. (\ref{Vbrij1}))
\Big($V^{br,\, extr2}_{LR\,(RL)\,ij}$ (see eq.
(\ref{Vbrij2}))\Big), as the ones coming from the calculations for
the positions $y_p\neq 0\,(y_p\neq 0,\, z_p\neq 0)$ of the 4D
brane in one (two) extra dimenion(s), over the ones, which are
obtained for the positions $y_p= 0\,(y_p=0,\, z_p= 0)$ of the 4D
brane, namely $V^{br,0}_{LR\,(RL)\,ij}$ (see eq. (\ref{V0brij1}))
\Big($V^{br,0,(2)}_{LR\,(RL)\,ij}$ (see eq. (\ref{V0brij2}))\Big),
included in the original Yukawa coupling $\xi^{E}_{ij}$ (see eq.
(\ref{coupbrl4}) \Big(eq. (\ref{coupbr2l44})\Big). On the other
hand, if the new Higgs doublet lies in one (two) extra
dimension(s) but it is restricted into the thin bulk, the
additional effects, $V^{bulk,extr}_{LR\,(RL)\,ij}$-
$V^{bulk,(n),extr}_{LR\,(RL)\,ij}$ (see eq.
(\ref{Vbuij1exa}-\ref{Vbunij1exa}))
\Big($V^{bulk,extr2}_{LR\,(RL)\,ij}$-$V^{bulk,
(n,s),extr2}_{LR\,(RL)\,ij}$ (see eq.
(\ref{Vbuij1a2}-\ref{Vbuij1a2ex}))\Big), are the ones coming from
the calculations for $w\leq 2\,\pi$ ($w_y\leq 2\,\pi$, $w_z\leq
2\,\pi$) in the bulk, over the ones, which are obtained for the
$w=2\,\pi$ ($w_y= 2\,\pi$, $w_z=2\,\pi$), in zero-KK mode level.
We include the zero level results, $V^{bulk,0}_{LR\,(RL)\,ij}$
(see eq. (\ref{V0buij1})) \Big($V^{bulk,0,0}_{LR\,(RL)\,ij}$ (see
eq. (\ref{V0buij2a}))\Big) in the original Yukawa coupling,
$\xi^{E}_{N,ij}$ (see eq. (\ref{coupbul4}) \Big(eq.
(\ref{coupbul44a2})\Big).

In our calculations, we take split leptons in a single and two
extra dimensions and use a possible set of locations to calculate
the contributions under consideration.  In the case of a single
extra dimension (two extra dimensions) we use the estimated
locations of the leptons given in eq. (\ref{location}) (eq.
(\ref{location2})) to calculate the lepton-lepton-Higgs scalar
vertices. For the parameter $\rho=\sigma/R$, where $\sigma$ is the
Gaussian width of the fermions (see \cite{Mirabelli} for details),
we use the numerical value $\rho=0.001$. Furthermore, we choose
the appropriate numerical values for the Yukawa couplings, by
respecting the current experimental measurements of these decays
(see Introduction section) and the muon anomalous magnetic moment
(see \cite{Iltananomuon} and references therein). Notice that, for
the Yukawa coupling $\bar{\xi}^{E}_{N,\tau \tau}$, we use the
numerical value which is greater than the upper limit of
$\bar{\xi}^{E}_{N,\tau \mu}$. The compactification scale $1/R$ is
another free parameter of the model and there are numerous
constraints for a single extra dimension in the split fermion
scenario. The direct limits from searching for KK gauge bosons
imply $1/R> 800\,\, GeV$, the precision electro weak bounds on
higher dimensional operators generated by KK exchange place a far
more stringent limit $1/R> 3.0\,\, TeV$ \cite{Rizzo} and from
$B\rightarrow \phi \, K_S$ the lower bounds for the scale $1/R$
have been obtained as $1/R > 1.0 \,\, TeV$, from $B\rightarrow
\psi \, K_S$ one got $1/R > 500\,\, GeV$, and from the upper limit
of the $BR$, $BR \, (B_s \rightarrow \mu^+ \mu^-)< 2.6\,\times
10^{-6}$, the estimated limit was $1/R > 800\,\, GeV$
\cite{Hewett}. We make our analysis by choosing two different
values of the compactification scale $1/R$, by respecting these
limits in the case of a single extra dimension. For two extra
dimensions, we used the same values of the scale $1/R$.

\begin{figure}[htb]
\vskip -3.0truein \centering \epsfxsize=6.8in
\leavevmode\epsffile{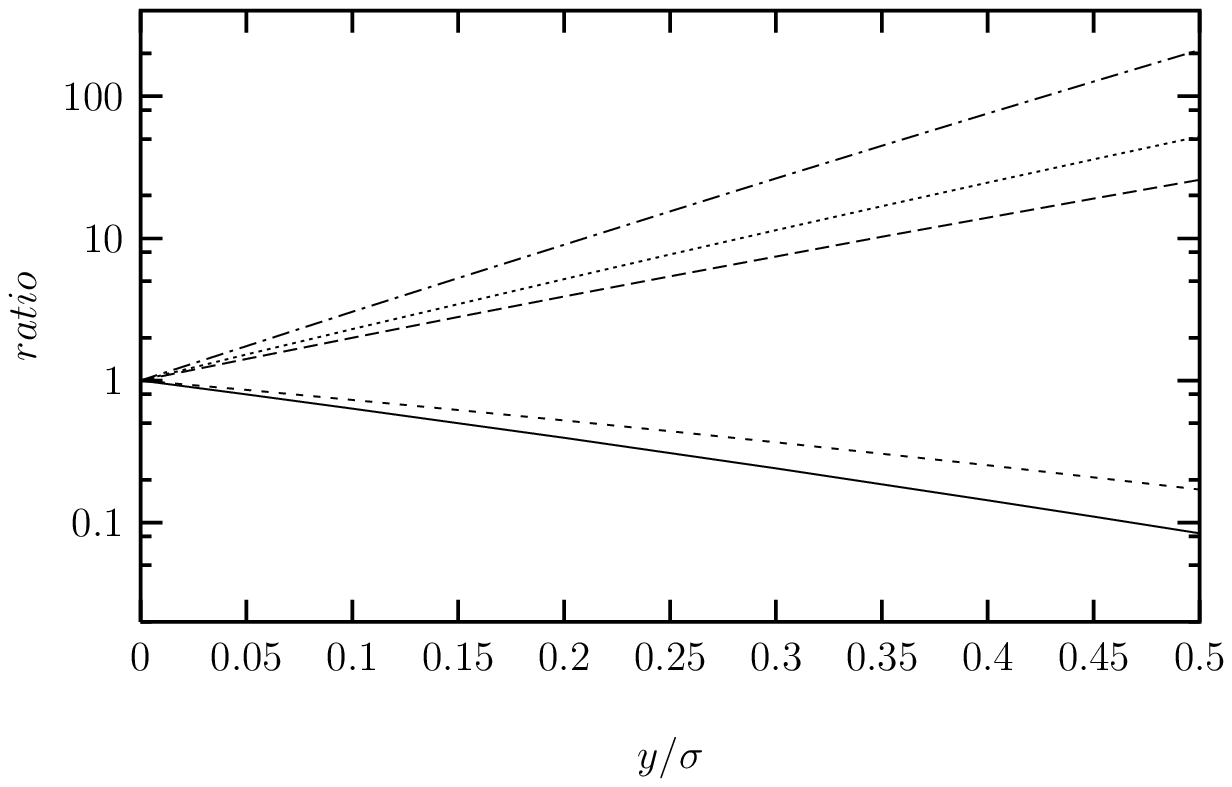} \vskip -3.0truein
\caption[]{$V^{br,extr}_{LR\,(RL)\,ij}$ with respect to
$y_p=y/\sigma$. Here the solid (dashed, small dashed, dotted,
dot-dashed) line represents the ratio for
$V^{br,extr}_{LR\,(RL)\,\tau\tau}$ ($V^{br,extr}_{LR\,\tau\mu}$,
$V^{br,extr}_{RL\,\tau\mu}$, $V^{br,extr}_{LR\,\tau e}$,
$V^{br,extr}_{RL\,\tau e}$).} \label{ratioonbraneex1}
\end{figure}

In  Fig. \ref{ratioonbraneex1}, we plot the ratio
$V^{br,extr}_{LR\,(RL)\,ij}=\frac{V^{br}_{LR\,(RL)\,ij}}
{V^{br,0}_{LR\,(RL)\,ij}}$ with respect to $y_p=y/\sigma$, which
is the location of the 4D brane, including the new Higgs bosons,
in the single extra dimension. Here the solid (dashed, small
dashed, dotted, dot-dashed) line represents the ratio for
$V^{br,extr}_{LR\,(RL)\,\tau\tau}$ ($V^{br,extr}_{LR\,\tau\mu}$,
$V^{br,extr}_{RL\,\tau\mu}$, $V^{br,extr}_{LR\,\tau e}$,
$V^{br,extr}_{RL\,\tau e}$). This figure shows that the ratios
$V^{br,extr}_{LR\,(RL)\,ij}$ (therefore, the effective Yukawa
couplings $V^{br,extr}_{LR\,(RL)\,ij}\,\bar{\xi}^E_{N,ij}$) are
sensitive the location $y_p$ of the 4D brane. The ratios for
$\tau_{L(R)} \tau_{R(L)} S$ and $\tau_{R} \mu_{L} S$ interactions
decrease and the others increase with the increasing values of the
location $y_p$. The enhancement in the values of the ratios for
the light flavors increases more compared to the heavy ones.
Respecting that the couplings $\bar{\xi}^{E}_{N,ij}$ with heavier
flavors are larger compared to lighter ones and with the similar
assumption for the effective Yukawa couplings, the location $y_p$
of the 4D brane, containing the new Higgs doublet should be in the
possible range, $0\leq y_p \leq \sim 0.3$. This is an interesting
result since the considered brane should be at the origin or near
to that point for the chosen free parameters and the location of
the lepton Gaussian profiles in the extra dimension, if the above
scenario is possible.

\begin{figure}[htb]
\vskip -3.0truein \centering \epsfxsize=6.8in
\leavevmode\epsffile{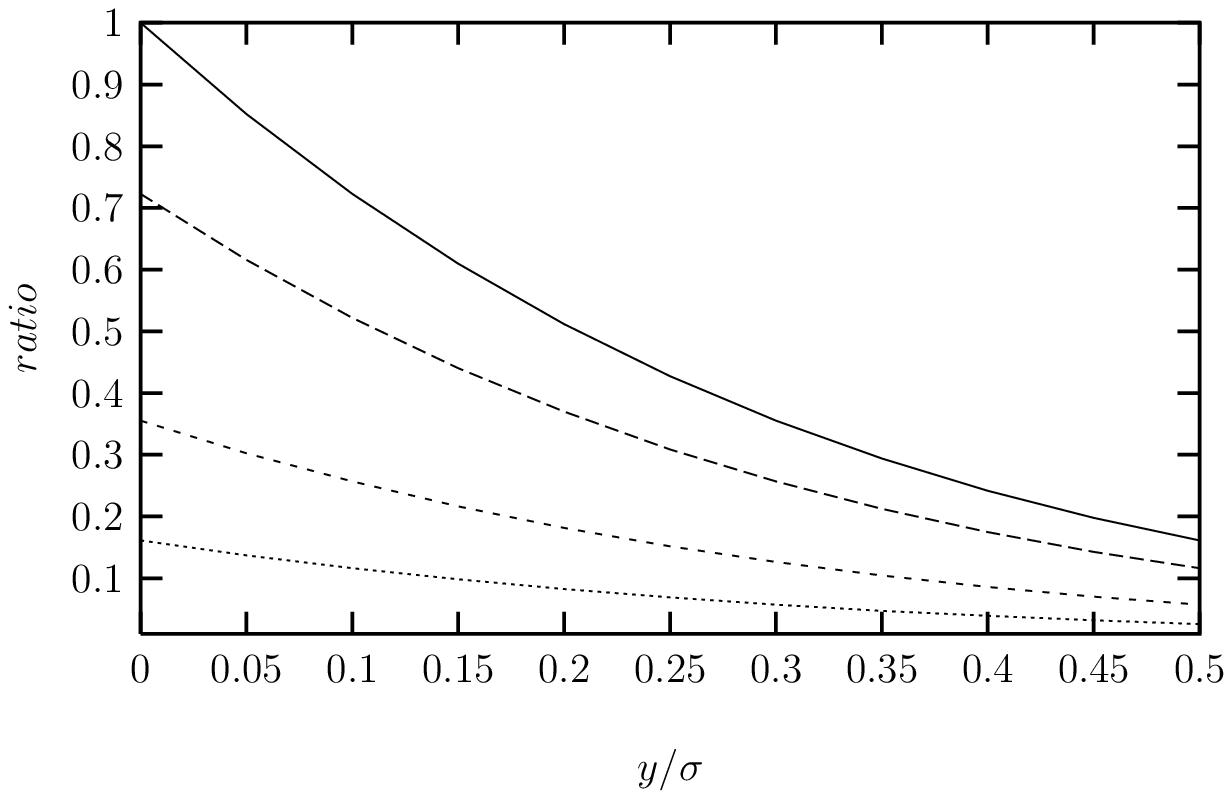} \vskip -3.0truein
\caption[]{$V^{br, extr2}_{LR\,(RL)\,\tau\tau}$ with respect to
$y_p=y/\sigma$, for four different values of the $z_p$. Here solid
(dashed, small dashed, dotted) line represents the ratios for
$z_p=0 (0.1, 0.3, 0.5)$.} \label{ratioonbranetauex2}
\end{figure}

\begin{figure}[htb]
\vskip -3.0truein \centering \epsfxsize=6.8in
\leavevmode\epsffile{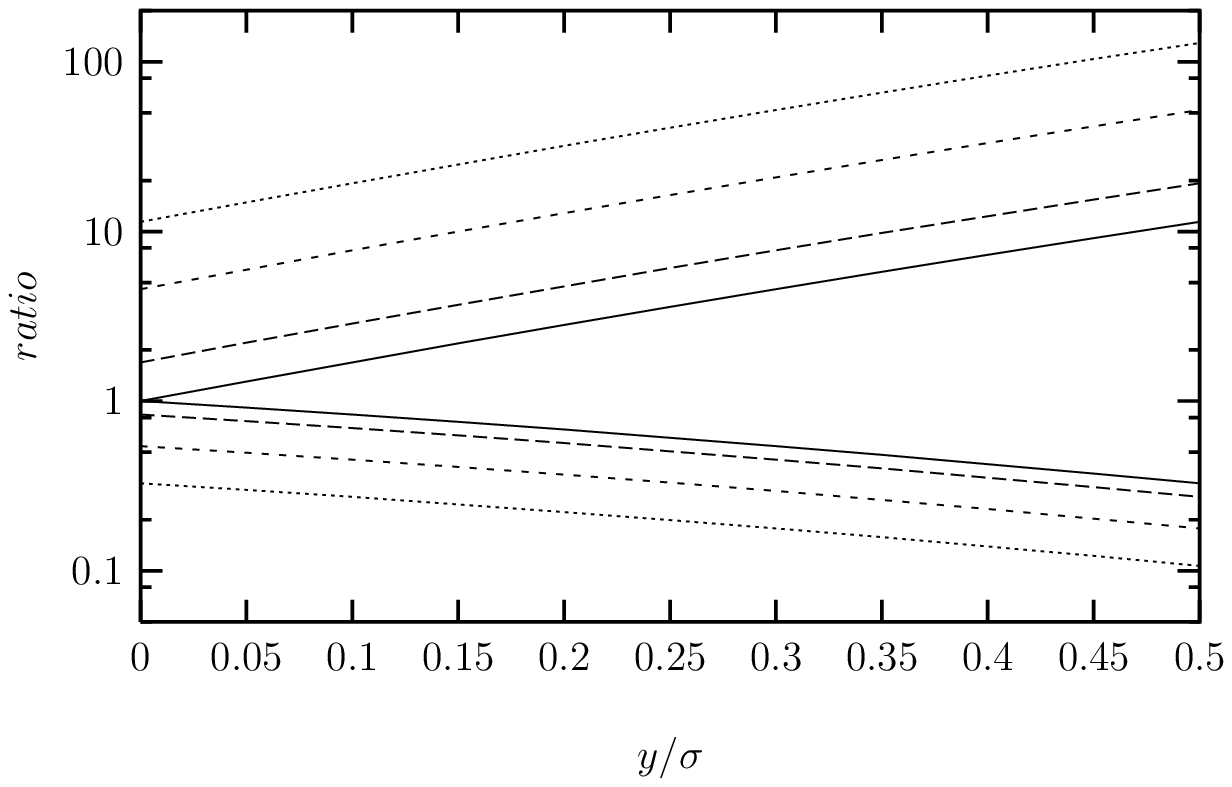} \vskip -3.0truein
\caption[]{The same as Fig. \ref{ratioonbranetauex2} but for
$V^{br, extr2}_{LR\,(RL)\,\tau\mu}$.} \label{ratioonbranemuex2}
\end{figure}

\begin{figure}[htb]
\vskip -3.0truein \centering \epsfxsize=6.8in
\leavevmode\epsffile{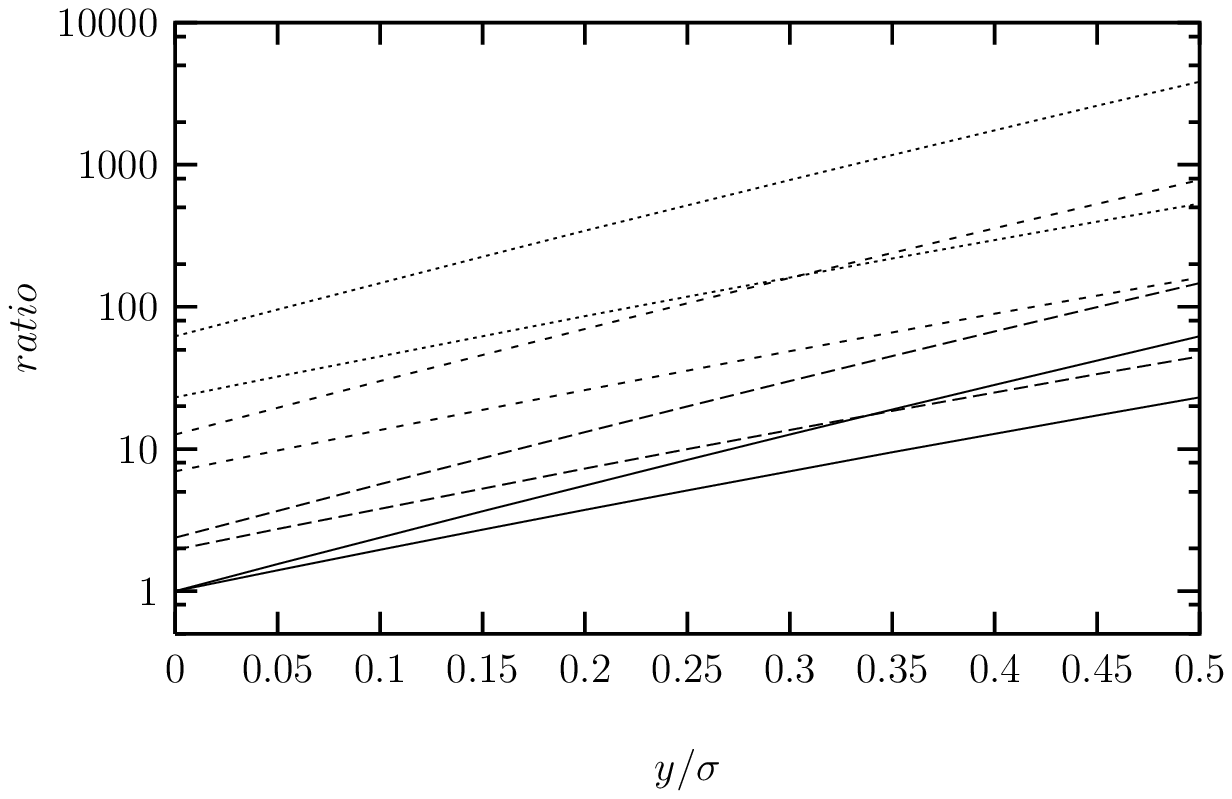} \vskip -3.0truein
\caption[]{The same as Fig. \ref{ratioonbranetauex2} but for
$V^{br, extr2}_{LR\,(RL)\,\tau e}$.} \label{ratioonbraneeex2}
\end{figure}

Fig. \ref{ratioonbranetauex2} represents the ratio $V^{br,
extr2}_{LR\,(RL)\,\tau\tau}=
\frac{V^{br,(2)}_{LR\,(RL)\,\tau\tau}}
{V^{br,0,(2)}_{LR\,(RL)\,\tau\tau}}$ with respect to
$y_p=y/\sigma$, for four different values of the $z_p=z/\sigma$,
which is the location of the 4D brane in the second extra
dimension. Here solid (dashed, small dashed, dotted) line
represents the ratios for $z_p=0\, (0.1, 0.3, 0.5)$. It is shown
that the ratio decreases with the increasing values of the
locations $y_p$ and also $z_p$. In Fig. \ref{ratioonbranemuex2}
(\ref{ratioonbraneeex2}) we present the ratio $V^{br,
extr2}_{LR\,(RL)\,\tau\mu}$ ($V^{br, extr2}_{LR\,(RL)\,\tau e}$)
with respect to $y_p=y/\sigma$, for four different values of the
$z_p$. Here the upper-lower solid (dashed, small dashed, dotted)
lines represent the ratios $V^{br, extr2}_{LR\,\tau\mu}$-$V^{br,
extr2}_{RL\,\tau\mu}$ and $V^{br, extr2}_{LR\,\tau e}$-$V^{br,
extr2}_{RL\,\tau e}$ for $z_p=0\, (0.1, 0.3, 0.5)$, in Fig.
\ref{ratioonbranemuex2} and \ref{ratioonbraneeex2}, respectively.
Fig. \ref{ratioonbranemuex2} shows that the ratios $V^{br,
extr2}_{LR\,\tau\mu}$ increase, while $V^{br,
extr2}_{RL\,\tau\mu}$ decrease with the increasing values of the
location $y_p$ and also $z_p$. On the other hand, both the ratios
$V^{br, extr2}_{LR\,\tau e}$ and $V^{br, extr2}_{RL\,\tau e}$ are
extremely sensitive to the location of the 4D brane in two extra
dimensions and increase with the increasing values of the
locations $y_p$ and $z_p$, especially for $V^{br, extr2}_{LR\,\tau
e}$. These enhancements switch on the increase in the values of
the effective couplings for the light flavors and, again,
respecting that the couplings $\bar{\xi}^{E}_{N,ij}$ and the
effective Yukawa couplings with heavier flavors are larger
compared to lighter ones, the location $y_p, z_p$ of the 4D brane,
containing the new Higgs doublet, should be in the possible range,
$0\leq y_p,\, z_p \leq \sim 0.2$. Similar to one extra dimension
case, the brane, including the new Higgs doublet, should be at the
origin or near to that point for the chosen free parameters and
lepton profiles.

\begin{figure}[htb]
\vskip -3.0truein \centering \epsfxsize=6.8in
\leavevmode\epsffile{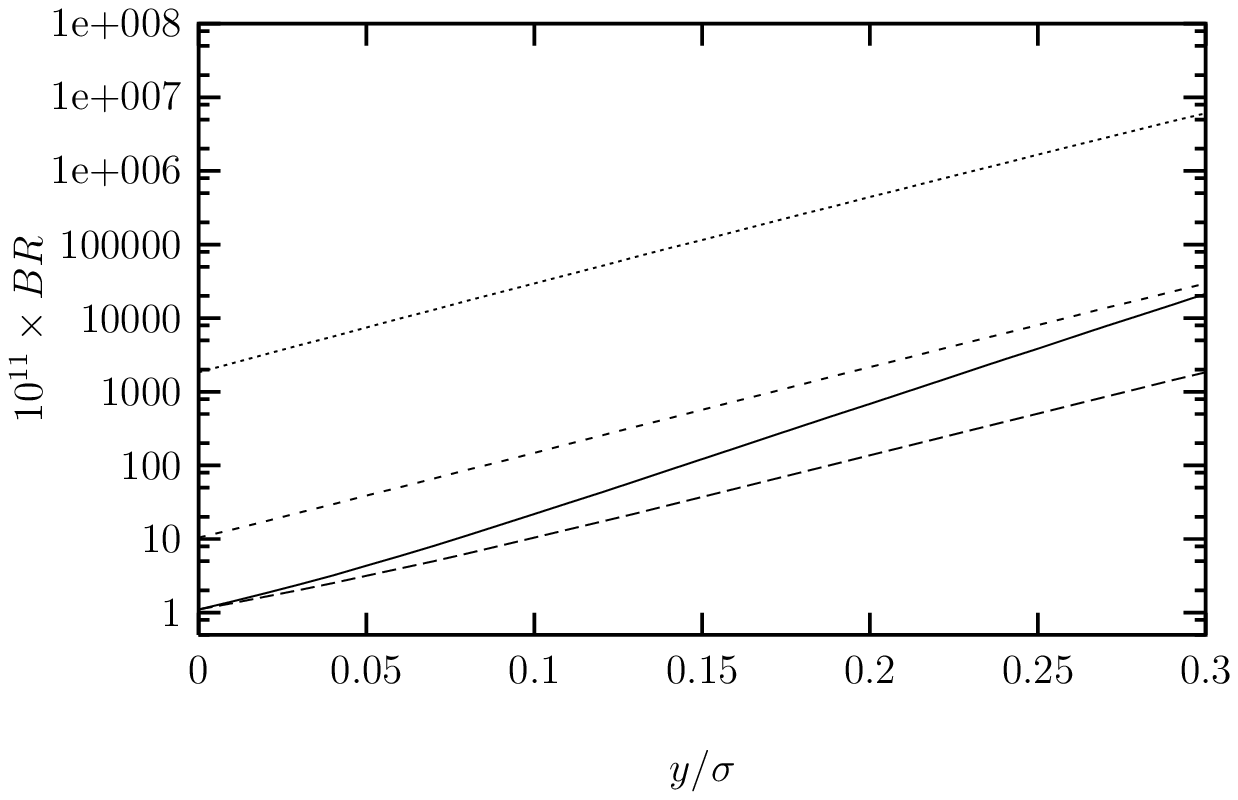} \vskip -3.0truein
\caption[]{BR ($\mu\rightarrow e \gamma$) with respect to the
$y_p=y/\sigma$, for $m_{h^0}=100\, GeV$, $m_{A^0}=200\, GeV$ and
the real couplings $\bar{\xi}^{E}_{N,\tau \mu} =10\, GeV$,
$\bar{\xi}^{E}_{N,\tau e} =0.001\, GeV$. Here the solid (dashed,
small dashed, dotted) line represents the BR for a single extra
dimension (for two extra dimensions and $z_p=0, 0.1, 0.3)$. }
\label{BRmuegamh0A0brane}
\end{figure}

Now, we would like to study the effects of the brane location on
the BRs of the LFV $l_i\rightarrow l_j \gamma$ decays. In  Fig.
\ref{BRmuegamh0A0brane}, we plot the BR of the decay
$\mu\rightarrow e \gamma$ with respect to the $y_p=y/\sigma$, for
$m_{h^0}=100\, GeV$, $m_{A^0}=200\, GeV$ and the real couplings
$\bar{\xi}^{E}_{N,\tau \mu} =10\, GeV$, $\bar{\xi}^{E}_{N,\tau e}
=0.001\, GeV$. Here the solid (dashed, small dashed, dotted) line
represents the BR for a single extra dimension (for two extra
dimensions and $z_p=0, 0.1, 0.3)$. It is observed that the BR is
strongly sensitive to the location of the 4D brane and increases
with the increasing values of $y_p$ ($y_p, z_p$). Not to exceed
the experimental value of the BR($\mu\rightarrow e\gamma$), namely
$1.2\times 10^{-11}$ \cite{Brooks}, one expects that the 4D brane,
including the new Higgs doublet, should be in the location $0\leq
y_p \leq \sim 0.05$ ($0\leq y_p,z_p \leq \sim 0.05$ in two
dimensions) \footnote{It is obvious that these numerical values
change with the different values of the couplings
$\bar{\xi}^{E}_{N,\tau \mu}$  and $\bar{\xi}^{E}_{N,\tau e}$.
However, the location of the 4D brane under consideration should
not be so far than the origin, if such scenario exists.}
\newpage

\begin{figure}[htb]
\vskip -3.0truein \centering \epsfxsize=6.8in
\leavevmode\epsffile{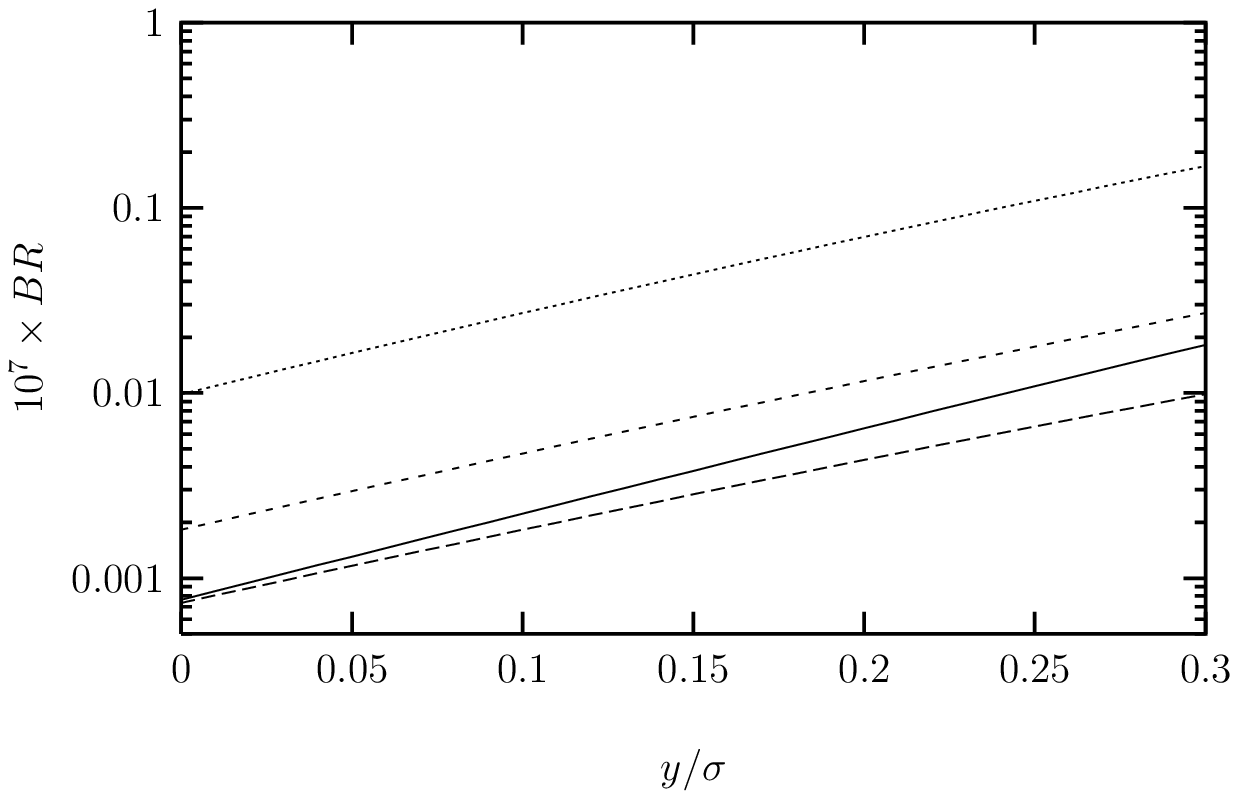} \vskip -3.0truein
\caption[]{ BR ($\tau\rightarrow e \gamma$) with respect to the
$y_p=y/\sigma$, for $m_{h^0}=100\, GeV$, $m_{A^0}=200\, GeV$ and
the real couplings $\bar{\xi}^{E}_{N,\tau \tau} =100\, GeV$,
$\bar{\xi}^{E}_{N,\tau e} =1\, GeV$. Here the solid (dashed, small
dashed, dotted) line represents the BR for a single extra
dimension (for two extra dimensions and $z_p=0, 0.1, 0.3)$. }
\label{BRtauegamh0A0brane}
\end{figure}

Fig. \ref{BRtauegamh0A0brane} is devoted to the BR of the decay
$\tau\rightarrow e \gamma$ with respect to the $y_p=y/\sigma$, for
$m_{h^0}=100\, GeV$, $m_{A^0}=200\, GeV$ and the real couplings
$\bar{\xi}^{E}_{N,\tau \tau} =100\, GeV$, $\bar{\xi}^{E}_{N,\tau
e} =1\, GeV$  \footnote{For $\tau\rightarrow e \gamma$ we take the
numerical value of the coupling $\bar{\xi}^{E}_{N,\tau e}$,
$\bar{\xi}^{E}_{N,\tau e} =1\, GeV$. Here we try to reach the new
experimental result of the BR of this decay (see \cite{Hayasaka}).
With the more sensitive future measurements of the BRs of these
decays these couplings would be fixed more accurately.}. Here the
solid (dashed, small dashed, dotted) line represents the BR for a
single extra dimension (for two extra dimensions and $z_p=0, 0.1,
0.3)$. This figure shows that the BR is sensitive to the location
of the 4D brane and increases with the increasing values of $y_p$
($y_p, z_p$). In this case, the BR ($\tau\rightarrow e\gamma$)
does not exceed the experimental value $3.9\times 10^{-7}$
\cite{Hayasaka} in the given region of the 4D brane, for the free
parameters used.

\begin{figure}[htb]
\vskip -3.0truein \centering \epsfxsize=6.8in
\leavevmode\epsffile{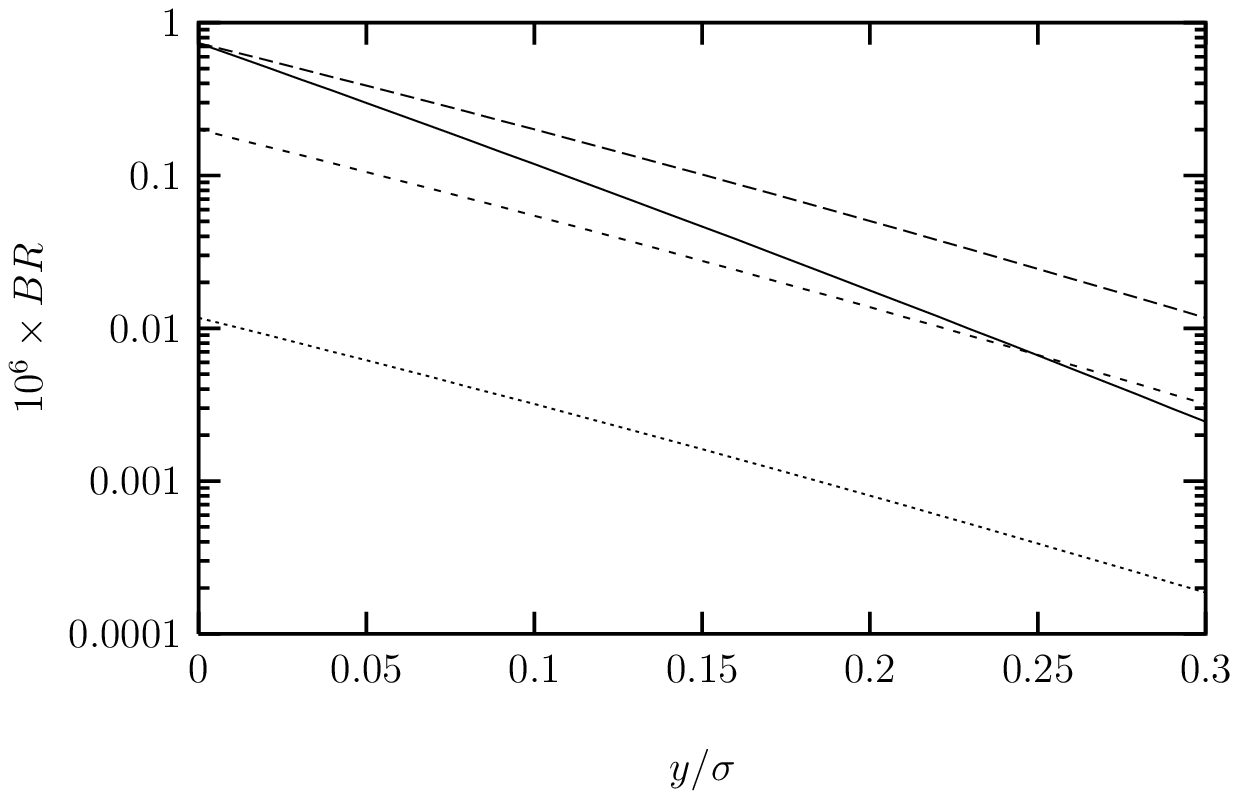} \vskip -3.0truein
\caption[]{ BR ($\tau\rightarrow \mu \gamma$) with respect to the
$y_p=y/\sigma$, for $m_{h^0}=100\, GeV$, $m_{A^0}=200\, GeV$ and
the real couplings $\bar{\xi}^{E}_{N,\tau \tau} =100\, GeV$,
$\bar{\xi}^{E}_{N,\tau \mu} =10\, GeV$. Here the solid (dashed,
small dashed, dotted) line represents the BR for a single extra
dimension (for two extra dimensions and $z_p=0, 0.1, 0.3)$.}
\label{BRtaumugamh0A0brane}
\end{figure}

Fig. \ref{BRtaumugamh0A0brane} represents the BR of the decay
$\tau\rightarrow \mu \gamma$ with respect to the $y_p=y/\sigma$,
for $m_{h^0}=100\, GeV$, $m_{A^0}=200\, GeV$ and the real
couplings $\bar{\xi}^{E}_{N,\tau \tau} =100\, GeV$,
$\bar{\xi}^{E}_{N,\tau \mu} =10\, GeV$. Here the solid (dashed,
small dashed, dotted) line represents the BR for a single extra
dimension (for two extra dimensions and $z_p=0, 0.1, 0.3)$. We see
that the BR is more sensitive to the location of the 4D brane
compared to the $\tau\rightarrow e \gamma$ decay and decreases
with the increasing values of $y_p$ ($y_p, z_p$). Not to be so
much below the experimental  value of the BR, $1.1\times 10^{-6}$
\cite{Ahmed} (also the recent upper limit of $9.0\, (6.8)\,
10^{-8}$ at $90\% CL$ \cite{Roney} (\cite{Aubert})), the 4D brane
including the new Higgs doublet should be near to the origin.
\footnote{With the more accurate future experimental measurements,
the upper limit can be pulled to the smaller values and one could
obtain more stringent restriction for the location of the 4D
brane.}

As another possibility, we assume that the new Higgs doublet in
the extra dimension(s) is localized into the thin bulk and we try
to examine the the effects of the width of the thin bulk to the
BRs of the LFV $l_i\rightarrow l_j \gamma$ decays.

\begin{figure}[htb]
\vskip -3.0truein \centering \epsfxsize=6.8in
\leavevmode\epsffile{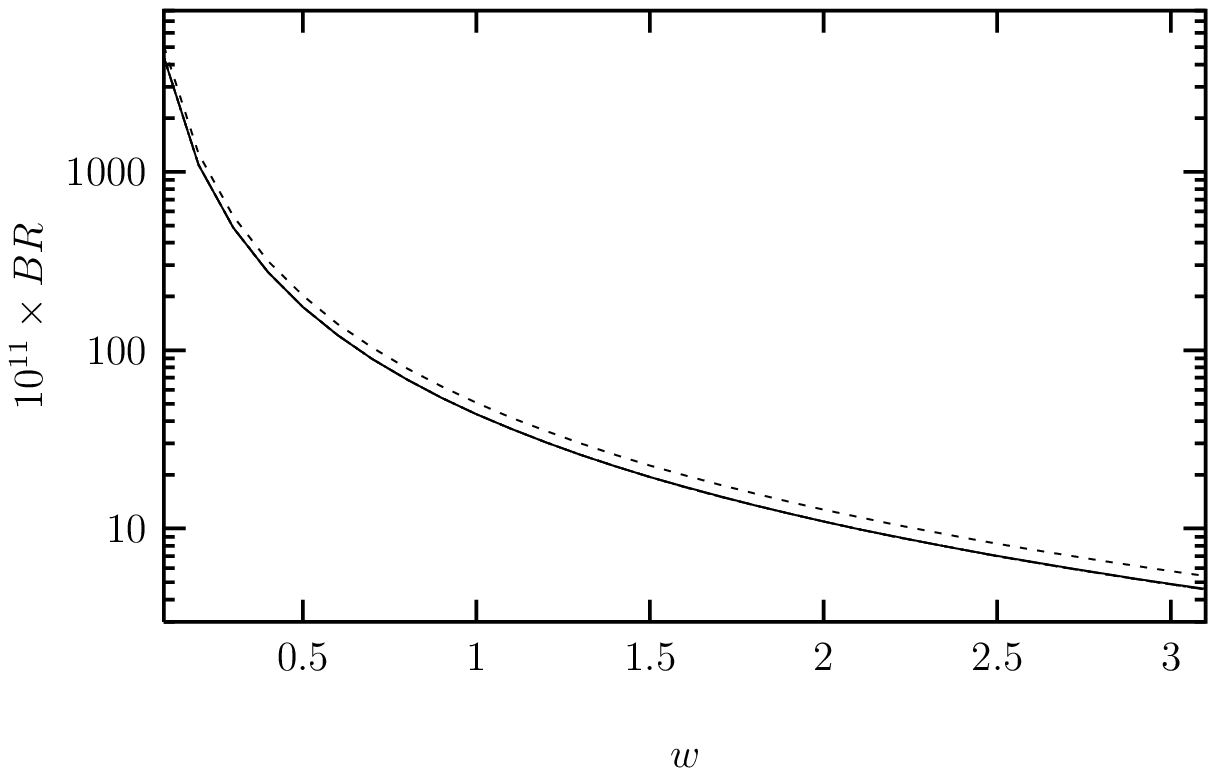} \vskip -3.0truein
\caption[]{ BR($\mu\rightarrow e \gamma$) with respect to the
parameter $w=width/R$, for $m_{h^0}=100\, GeV$, $m_{A^0}=200\,
GeV$ and the real couplings $\bar{\xi}^{E}_{N,\tau \mu} =10\,
GeV$, $\bar{\xi}^{E}_{N,\tau e} =0.001\, GeV$. Here the solid
(dashed) line represents the BR for one-two extra dimensions and
$1/R=5000\,GeV$ (for two extra dimensions and $1/R=500\,GeV$).}
\label{BRmuegamAllBulk}
\end{figure}

Fig. \ref{BRmuegamAllBulk} is devoted to the BR of the decay
$\mu\rightarrow e \gamma$ with respect to the the parameter
$w=width/R$, for $m_{h^0}=100\, GeV$, $m_{A^0}=200\, GeV$ and the
real couplings $\bar{\xi}^{E}_{N,\tau \mu} =10\, GeV$,
$\bar{\xi}^{E}_{N,\tau e} =0.001\, GeV$. Here the solid (dashed)
line represents the BR for one-two extra dimensions and
$1/R=5000\,GeV$ (for two extra dimensions and $1/R=500\,GeV$). It
is observed that the BR is strongly sensitive to the width of the
thin bulk and it should be $w>3$, not to exceed  the experimental
value of the BR.

\begin{figure}[htb]
\vskip -3.0truein \centering \epsfxsize=6.8in
\leavevmode\epsffile{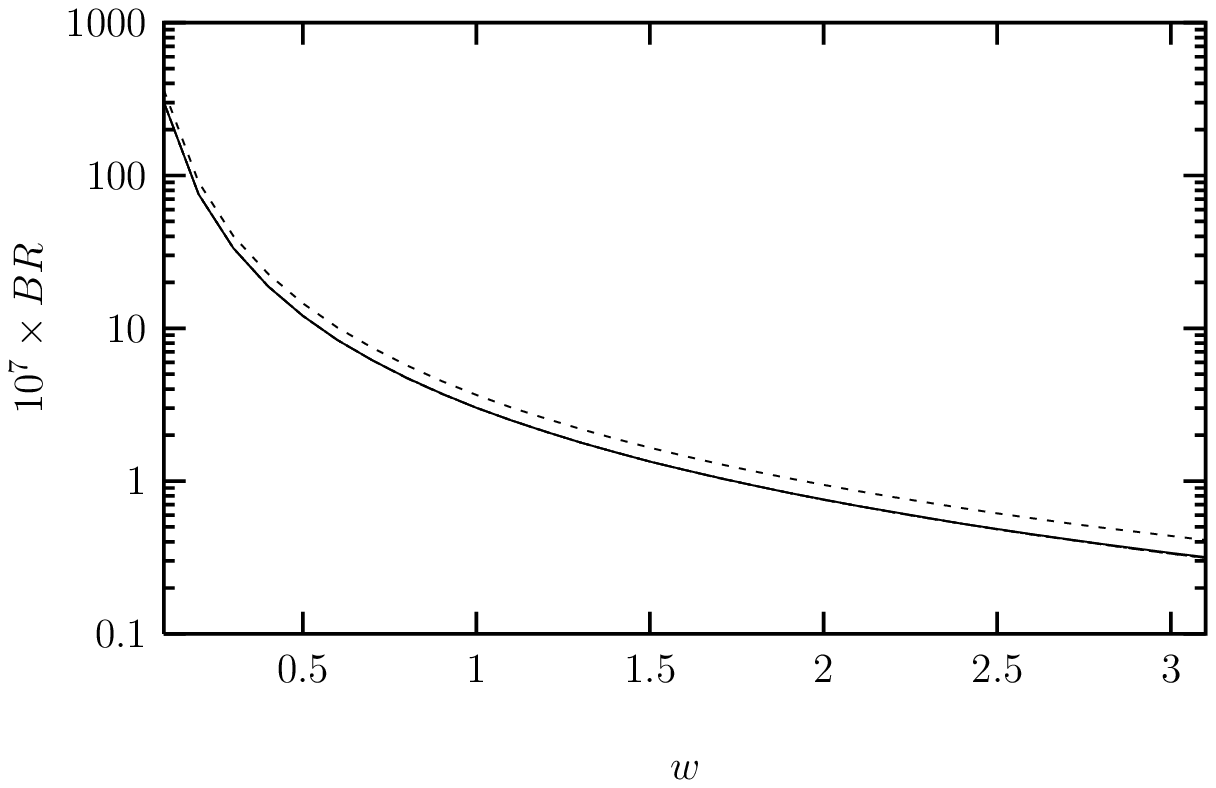} \vskip -3.0truein
\caption[]{ BR($\tau\rightarrow e \gamma$) with respect to the
parameter $w=width/R$, for $m_{h^0}=100\, GeV$, $m_{A^0}=200\,
GeV$ and the real couplings $\bar{\xi}^{E}_{N,\tau \tau} =100\,
GeV$, $\bar{\xi}^{E}_{N,\tau e} =1\, GeV$. Here the solid (dashed)
line represents the BR for one-two extra dimensions and
$1/R=5000\,GeV$ (for two extra dimensions and $1/R=500\,GeV$).}
\label{BRtauegamAllBulk}
\end{figure}

In Fig. \ref{BRtauegamAllBulk} we present the BR of the decay
$\tau\rightarrow e \gamma$ with respect to the the parameter
$w=width/R$, for $m_{h^0}=100\, GeV$, $m_{A^0}=200\, GeV$ and the
real couplings $\bar{\xi}^{E}_{N,\tau \tau} =100\, GeV$,
$\bar{\xi}^{E}_{N,\tau e} =1\, GeV$. Here the solid (dashed) line
represents the BR for one-two extra dimensions and $1/R=5000\,GeV$
(for two extra dimensions and $1/R=500\,GeV$). We see that the BR
is sensitive to the width of the thin bulk and the width of the
thin bulk should be $w>2$, not to exceed the experimental value of
the BR of the decay considered.

\begin{figure}[htb]
\vskip -3.0truein \centering \epsfxsize=6.8in
\leavevmode\epsffile{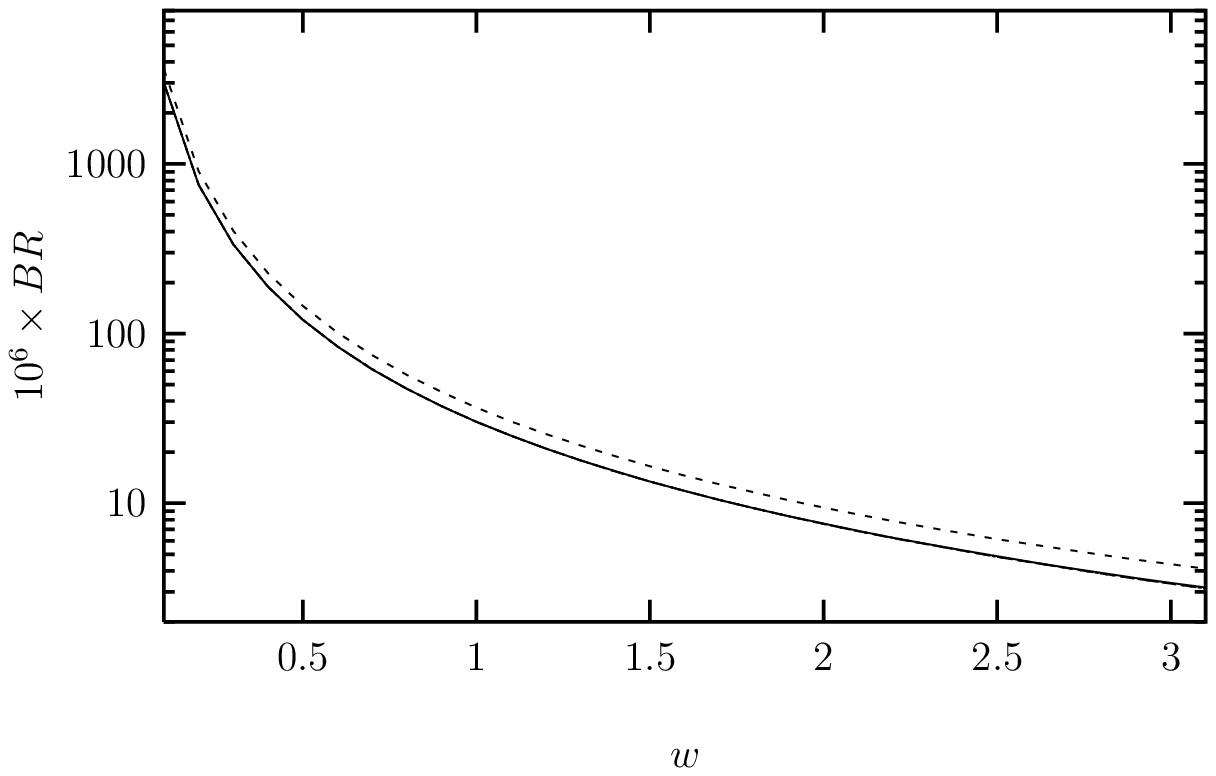} \vskip -3.0truein
\caption[]{ BR($\tau\rightarrow \mu \gamma$) with respect to the
parameter $w=width/R$, for $m_{h^0}=100\, GeV$, $m_{A^0}=200\,
GeV$ and the real couplings $\bar{\xi}^{E}_{N,\tau \tau} =100\,
GeV$, $\bar{\xi}^{E}_{N,\tau \mu} =10\, GeV$. Here the solid
(dashed) line represents the BR for one-two extra dimensions and
$1/R=5000\,GeV$ (for two extra dimensions and $1/R=500\,GeV$).}
\label{BRtaumugamAllBulk}
\end{figure}

Fig. \ref{BRtaumugamAllBulk} represents the BR of the decay
$\tau\rightarrow \mu \gamma$ with respect to the the parameter
$w=width/R$, for $m_{h^0}=100\, GeV$, $m_{A^0}=200\, GeV$ and the
real couplings $\bar{\xi}^{E}_{N,\tau \tau} =100\, GeV$,
$\bar{\xi}^{E}_{N,\tau \mu} =10\, GeV$. Here the solid (dashed)
line represents the BR for one-two extra dimensions and
$1/R=5000\,GeV$ (for two extra dimensions and $1/R=500\,GeV$). We
see that the BR is strongly sensitive to the width of the thin
bulk, similar to the $\mu\rightarrow e \gamma$ decay and the width
of the thin bulk should be $w>3$.

At this stage we would like to summarize our results:
\begin{itemize}
\item  If the new Higgs doublet is located in the 4D brane, in the
extra dimension(s), the BRs of LFV $l_i\rightarrow l_j\gamma$
decays are strongly sensitive to the location of the 4D brane. For
$\mu\rightarrow e\gamma$ and $\tau\rightarrow e\gamma$
($\tau\rightarrow \mu\gamma$) decays the BRs increase (decrease)
with the increasing values of $y_p$ and, in two extra dimensions,
$y_p, z_p$. The considered 4D brane  should be located near to the
origin in the extra dimension(s) not to have a conflict with the
current experimental results of the BRs.
\item If we assume that the new Higgs doublet in the extra
dimension(s) is localized into the thin bulk, it is estimated that
the BR is strongly sensitive to the width of the thin bulk and its
width should be approximately $w>3$, not to exceed  the current
experimental results of the BRs of LFV $l_i\rightarrow l_j\gamma$
decays.
\end{itemize}

With the help of the forthcoming most accurate experimental
measurements of the radiative LFV decays, a considerable
information can be obtained to restrict the free parameters and to
check the split fermion scenarios with the new Higgs sector.
\section{Acknowledgement}
This work has been supported by the Turkish Academy of Sciences in
the framework of the Young Scientist Award Program.
(EOI-TUBA-GEBIP/2001-1-8)
\newpage
\newpage
\end{document}